\documentclass{PoS}

\usepackage{amsmath}
\usepackage{amsfonts}
\usepackage{amssymb}
\usepackage{latexsym}
\usepackage{graphicx}
\usepackage{float}

\usepackage{algorithm,setspace}
\usepackage{algpseudocode}

\usepackage{multirow}

\usepackage[dvipsnames]{xcolor}

\usepackage{colortbl}
\definecolor{LightCyan}{rgb}{0.88,1,1}

\usepackage{mathtools}

\title{Solving DWF Dirac Equation Using Multisplitting Preconditioned Conjugate Gradient}

\ShortTitle{}

\author{\speaker{Jiqun Tu}\thanks{The speaker wishes to thank his advisor Robert Mawhinney for the support for this work.  The author also wants to express gratitude to Guo Duo, Chulwoo Jung, Christopher Kelly and Norman Christ for the suggestions and comments. The algorithm is implemented numerically with the help from these libraries: \texttt{CPS}, {{\texttt{Grid}}}, {\texttt{Qlattice}} and {\texttt{QUDA}}. Thanks to Kate Clark and NVIDIA for the numerical support.}\\
        Department of Physics, Columbia University, New York, NY 10027, USA\\
        E-mail: \email{jt2798@columbia.edu}}

\abstract{We show that using the multisplitting algorithm as a preconditioner for conjugate gradient inver- sion of the domain wall fermion Dirac operator could effectively reduce the inter-node communication cost, at the expense of performing more on-node floating point operations. This method could be useful for supercomputers with far more on-node flops than inter-node communication bandwidth.}

\FullConference{The 36th Annual International Symposium on Lattice Field Theory - LATTICE2018\\
		22-28 July, 2018\\
		Michigan State University, East Lansing, Michigan, USA.}

\usepackage[style=phys,articletitle=false,biblabel=brackets,chaptertitle=false,pageranges=false,maxbibnames=3,url=false,doi=false,eprint=true,isbn=false]{biblatex}
\DeclareFieldFormat{eprint:arxiv}{%
  	\iffieldundef{eprintclass}{%
  			\mkbibbrackets{\href{http://arxiv.org/abs/#1}{\texttt{arXiv:#1}}}
  	}{%
  		\mkbibbrackets{%
    		\href{http://arxiv.org/abs/\thefield{eprintclass}/#1}{\texttt{arXiv:\thefield{eprintclass}/#1}}}
	}
}

\DeclareFieldFormat{doi/url-link}{%
      {#1}%
}

\DeclareBibliographyDriver{article}{%
  \usebibmacro{bibindex}%
  \usebibmacro{begentry}%
  \usebibmacro{author/translator+others}%
  \setunit{\labelnamepunct}\newblock
  \usebibmacro{title}%
  \newunit
  \printlist{language}%
  \newunit\newblock
  \usebibmacro{byauthor}%
  \newunit\newblock
  \usebibmacro{bytranslator+others}%
  \newunit\newblock
  \printfield{version}%
  \newunit\newblock
  \printtext[doi/url-link]{%
    \usebibmacro{journal+issuetitle}%
    \newunit
    \usebibmacro{byeditor+others}%
    \newunit
    \usebibmacro{note+pages}%
    \newunit\newblock
    \iftoggle{bbx:isbn}
      {\printfield{issn}}
      {}%
    \setunit{\addspace}%
    \printfield{year}%
  }%
  \setunit{\addspace}%
  \iffieldundef{pages}
    {%
    }%
    {}%
  \usebibmacro{doi+eprint+url}%
  \newunit\newblock
  \usebibmacro{addendum+pubstate}%
  \setunit{\bibpagerefpunct}\newblock
  \usebibmacro{pageref}%
  \newunit\newblock
  \usebibmacro{related}%
  \usebibmacro{finentry}%
}

\begin{document}

\section{Introduction}

The cost of lattice QCD simulations with dynamical fermions is dominated by the solution of the Dirac equation in both the ensemble generation phase, where configurations of gauge fields are generated, and the measurement phase, where expectation values of physical observables are measured. 
The Dirac matrix, which is the gauge field dependent discretization of the fermionic part of the continuous QCD action, is a large sparse linear system and inverting the corresponding Dirac equation poses tremendous numerical difficulty. 
For domain wall fermions(DWF) the conjugate gradient(CG) algorithm proves to be a stable algorithm to solve the Dirac equation but the convergence rate is limited by the condition number of the Dirac matrix, which is typically large in simulations with physical pion mass.  

For the measurement phase various eigen-space methods, including EigCG\cite{Stathopoulos2010} and implicitly restarted Lanczos algorithm with Chebyshev polynomial\cite{YSaad1980}, have been developed successfully to speed up the inversion.
Low-lying eigenvectors(eigenvectors corresponding to small eigenvalues) of the Dirac matrix are generated and the previously large condition number is effectively reduced to improve the convergence rate of CG.
In this phase for one gauge field configuration typically a large number of Dirac equations with the same Dirac matrix but different right hand sides(RHS, or sources) are solved. The large number of sources amortizes the cost of eigenvector generation and the total computation time is reduced.

This is not the case for the ensemble generation phase. During a typical hybrid Monte Carlo(HMC) evolution of a gauge field as few as one Dirac equation is solved for a single Dirac matrix. This renders it not worthwhile to generate the low-lying eigenvectors for a particular Dirac matrix.

The development of supercomputers has greatly increased the number of floating point operations per second(flops) that can be performed on each processor(node).
Modern lattice simulations usually divide the gauge field and pseudo-fermion fields into sub-fields that are stored and computed locally on different processors of a large parallel computer.
This increases the total theoretical floating point operation capability.
Inter-processor data transfer(communication), however, is needed to perform coherent operations, including the Dirac matrix multiplication. Computations locally performed on one processor require contents of the sub-fields that are stored and updated on other processors.
For a specific operation if the rate of communication could not keep up with the local flops then communication becomes the bottleneck and the high flops are not utilized. 

For standard CG solver with DWF one Dirac matrix multiplication is performed for each iteration.  The precise requirement varies with the size of the lattice and processor grid, but roughly this requires one byte of communication for each local floating point operation.
On some of the newest machines, for example the SUMMIT machine at Oak Ridge National Laboratory(ORNL), inter-processor communication speed is much less than the  requirement set by their high local floating point operation capability.

In \cite{Luscher2004} a domain decomposition algorithm is proposed for Dirac equation with Wilson fermion. Local inversions are performed on two halves of the lattice iteratively. However, attempts to apply the same or similar algorithms to the inversion of the DWF Dirac equation have not been successful. 


In this work we report on our investigation into a preconditioned CG solver for solving the DWF Dirac equation for the ensemble generation phase of the simulation. We find a preconditioner that decreases the number of CG iterations needed for a solution, while increasing the local computation required per iteration, thus changing the balance of local computation to off-processor communication.


%

\section{Method}
\subsection{Multisplitting Algorithm}
In \cite{OLeary1985} a \textit{multisplitting} algorithm is proposed for solving generic large linear systems distributed across a parallel computer.
Compared to the domain decomposition algorithm in \cite{Luscher2004}, it does not require checkerboarding.
Before each iteration the boundary content of the solution field on each of the processors is communicated to its neighbors.
During each iteration, the algorithm uses this communicated neighboring solution field as the Dirichlet boundary condition to perform the inversion of a local matrix on each processor. 
After each iteration, the updated boundary content is again communicated to prepare for the next iteration.

\begin{figure}[]
	\centering
	\includegraphics[width=0.8\textwidth]{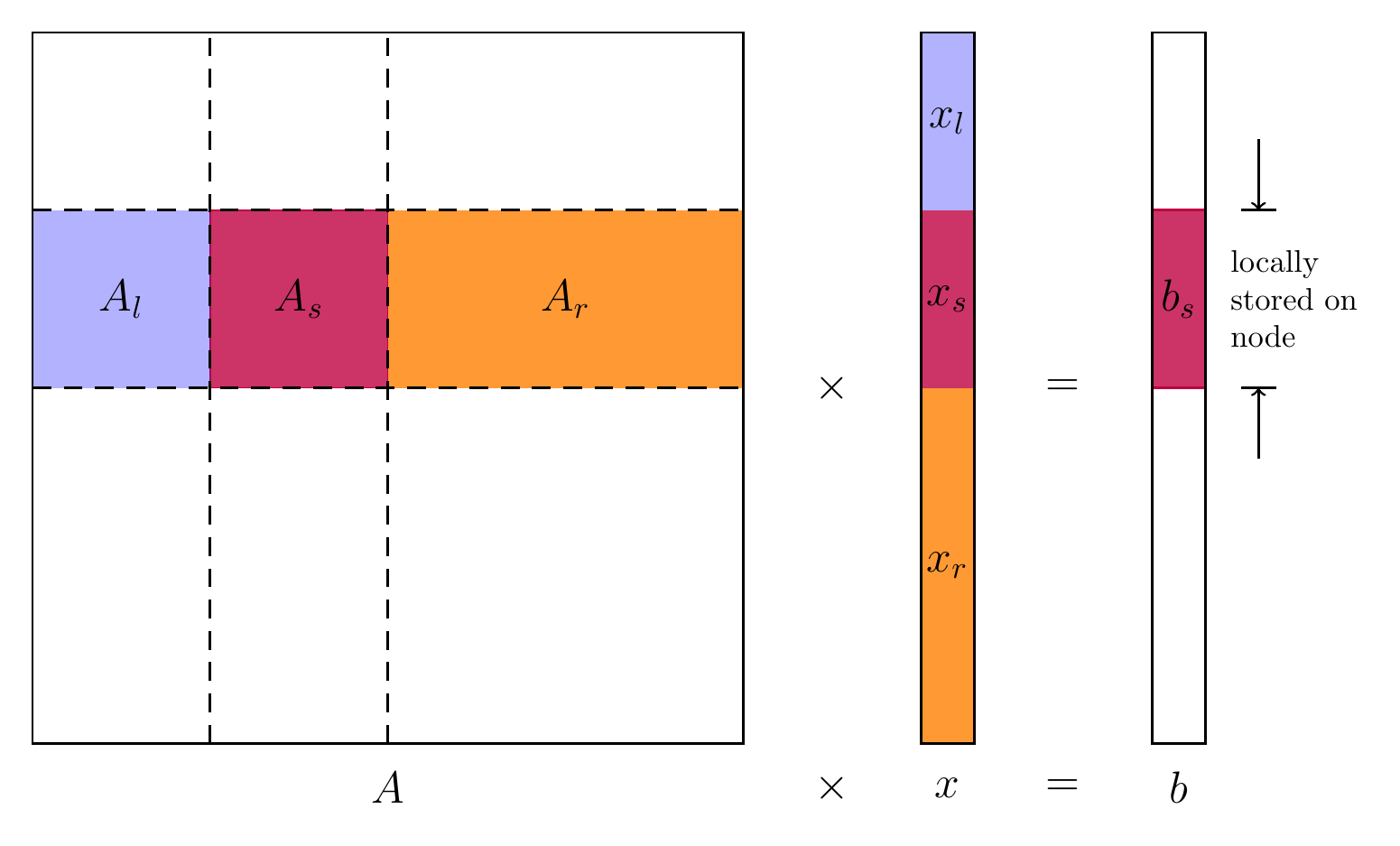}
	\caption{Decomposition of the matrix $A$, the solution vector $x$ and the right-hand-side(RHS) vector $b$ into local parts on each node.}\label{fig:ms_dec}
\end{figure}
Following \cite{Jezequel2012}, suppose the equation to be solved is $Ax=b$. For a \textit{particular} processor the matrix $A$ and vectors $x$ and $b$ are decomposed according to figure \ref{fig:ms_dec}, where $x_s$ and $b_s$ are the part that is locally stored on this processor. On each processor the original equation turns into
\begin{equation}\label{local}
	A_sx_s+A_lx_l+A_rx_r=b_s.
\end{equation} 
The $A_lx_l+A_rx_r$ part involves off-processor content and is calculated before each iteration via communication. $A_s$ is the part of the matrix that requires only the locally stored part of $x$ on a certain processor $s$, i.e. $x_s$. Then for each iteration the algorithm solves the equation 
\begin{equation}\label{eq:ms}
	A_sx_s=b_s-A_lx_l-A_rx_r
\end{equation}
locally for $x_s$ on this processor. The updated solution $x_s$ will then be communicated to the neighboring processors. This whole procedure can be done concurrently on all nodes once the communication work to calculate $A_lx_l+A_rx_r$ is done.


\subsection{Domain Wall Fermions}
The domain wall fermion(DWF)\cite{Jansen1996} formulation is based on Wilson fermion and a fictitious fifth dimension. Modern numerical implementations of DWF utilize the fact that only the matrix elements that connect the \textit{even} sites to \textit{odd} sites and those connecting \textit{odd} sites to \textit{even} sites depend the gauge field. The matrix entries that connect \textit{even} sites to \textit{even} sites and those connect \textit{odd} sites to \textit{odd} sites are constant. Here the even-odd parity is defined by the 4D components of a site:
\begin{equation}
\mathrm{parity}\equiv (x+y+z+t)\mod 2.  
\end{equation}
In the 4D even-odd preconditioning form the M\"obius DWF Dirac equation can be written as, 
\begin{equation}
 \begin{pmatrix}
  M_5 & M^4_{eo} \\
  M^4_{oe} & M_5 \\
 \end{pmatrix}
  \begin{pmatrix}
  \psi_e \\
  \psi_o
  \end{pmatrix}
  =
  \begin{pmatrix}
  \phi_e \\
  \phi_o
  \end{pmatrix},  
\end{equation}
where the subscript $e/o$ refer to even and odd sites. This is equivalent to solving the following even-odd preconditioned equation,
\begin{equation}\label{dirac_equation}
D_{PC}\psi_e=\hat\phi_e,\ D_{PC}\equiv M_5-M^4_{eo}M_5^{-1}M^4_{oe}, \hat\phi_e\equiv\phi_e-M^4_{eo}M_5^{-1}\phi_o. 
\end{equation}
Here $M^4_{eo/oe}$ includes the Wilson hopping term $D^w_{x,y}$ that connects 4D space-time sites to their nearest neighbors,
\begin{equation}
M^4_{oe/eo}=D^w_{x,y}M_\phi,\ D^w_{x,y}\equiv\sum_\mu\left[(1+\gamma_\mu)U^\dagger_{x-\hat{\mu},\mu}\delta_{x-\hat\mu,y}+(1-\gamma_\mu)U^\dagger_{x,\mu}\delta_{x+\hat{\mu},y}\right],
\end{equation}
and $M_5$ and $M_\phi$ are constant matrices that are diagonal in the four Euclidean space-time dimensions. Details of these matrices can be found in \cite{Brower2014}.

The CG algorithm requires the matrix to be hermitian and positive definite. A common practice is to multiply both sides of (\ref{dirac_equation}) with $D^\dagger_{PC}$ and solve the equation with the normal operator $D^\dagger_{PC}D_{PC}$ and the new RHS $D_{PC}^\dagger\hat\phi_e$ instead,
\begin{equation}\label{normal_equation}
D^\dagger_{PC}D_{PC}\psi_e=D_{PC}^\dagger\hat\phi_e.  
\end{equation}

\subsection{Dirichlet Boundary Condition on the 4-Hop Normal Operator}
There are four Wilson hopping terms, one in each $M_{eo/oe}^4$, in the normal operator $D^\dagger_{PC}D_{PC}$,
\begin{equation}\label{eq:DdagD}
  D^\dagger_{PC}D_{PC}=\big[M_5-\textcolor{red}{M^4_{eo}}M_5^{-1}\textcolor{red}{M^4_{oe}}\big]^\dagger\big[M_5-\textcolor{red}{M^4_{eo}}M_5^{-1}\textcolor{red}{M^4_{oe}}\big].
\end{equation}

\begin{figure}[]
	\centering
	\includegraphics[width=0.6\textwidth]{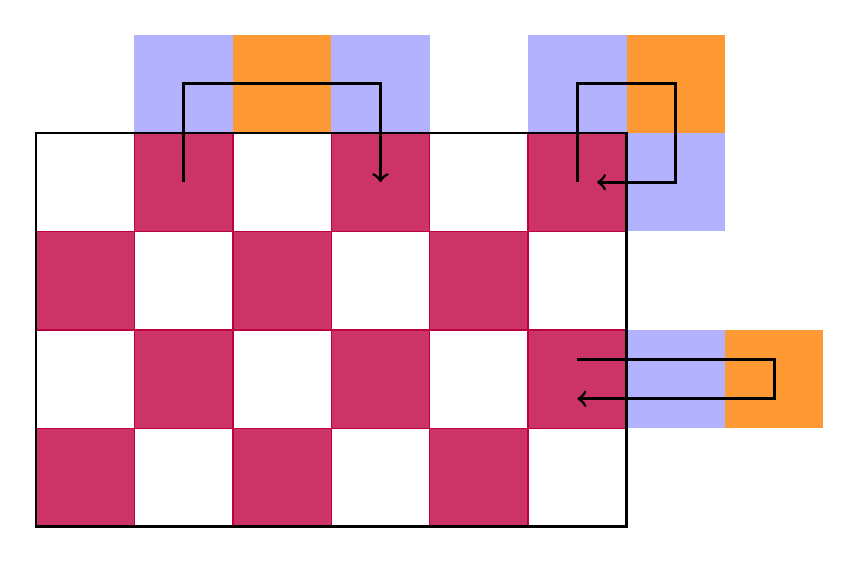}
	\caption{The normal operator $D^\dagger_{PC}D_{PC}$ has as many as $4$ Wilson hopping terms. Enforcing Dirichlet boundary condition on it requires the inclusion of the \textit{snake} terms, e.g. the black arrows.}\label{fig:snake}
\end{figure}

To apply the multisplitting algorithm to equation (\ref{normal_equation}) Dirichlet boundary conditions are to be enforced on the normal operator $D^\dagger_{PC}D_{PC}$, i.e. the local part(the $A_s$ in (\ref{local})) of this normal operator needs to be constructed. As the vector content is distributed across the processors according to its 4D space-time location, this local part for $D^\dagger_{PC}D_{PC}$ includes \textit{snake} terms that hop out of the boundary and hop back in as the various components in (\ref{eq:DdagD}) are evaluated. Figure \ref{fig:snake} illustrates this and gives some examples of the snake terms. These terms are truncated if Dirichlet boundary conditions are enforced on each of the four $M^4_{eo/oe}$ hopping terms sequentially. Our simulation results show that the inclusion of these snake terms is crucial to the convergence.

\subsection{Multisplitting Algorithm as a Preconditioner of CG}
In \cite{Luscher2004} to achieve faster convergence the domain decomposition algorithm is eventually used as a preconditioner of GCR. In this work we use the multisplitting algorithm as a preconditioner of CG.

Pseudocode for a generic preconditioned CG is shown below, where we are solving $Ax=b$ and $M$ is the preconditioning matrix. The preconditioning step is marked with blue background. The overall convergence rate of preconditioned CG is estimated by the condition number of $AM^{-1}$. If the condition number of $AM^{-1}$ is smaller then that of the original matrix $A$, faster convergence rate is achieved.
\begin{algorithm}
\setstretch{1.15}
\caption{Preconditioned Conjugate Gradient $Ax=b$}
\begin{algorithmic}
\State ${r}_0 = {b} - {A x}_0$
\State ${z}_0 = {M}^{-1} {r}_0$ 
\State ${p}_0 = {z}_0$ 
\State $k = 0$  
\While {have not converged}
\State $\alpha_k = {\langle{r}_k,{z}_k\rangle}/{\langle{p}_k,{A p}_k \rangle}$ 
\State ${x}_{k+1} = {x}_k + \alpha _k {p}_k$ 
\State ${r}_{k+1} = {r}_k - \alpha _k {A p}_k$ 
\State \colorbox{blue!30}{${z}_{k+1} = {M}^{-1} {r}_{k+1}$}
\State $\beta _k = {\langle {z}_{k+1}, {r}_{k+1}\rangle}/{\langle {z}_k,{r}_k \rangle}$ 
\State ${p}_{k+1} = {z}_{k+1} + \beta _k {p}_k$ 
\State $k = k + 1$ 
\EndWhile
\end{algorithmic}
\end{algorithm}

Now for this preconditioning step we use the multisplitting algorithm to solve for $z_{k+1}$ in
\begin{equation}
  Az_{k+1}=r_{k+1}.
\end{equation}
To avoid inter-processor communication, a zero initial guess($x_l=x_r=0$) is used in (\ref{eq:ms}) and only the first iteration is performed. With $r_{k+1}$ as the RHS and $z_{k+1}$ the solution,
\begin{equation}
  A_s x_s = b_s -A_lx_l - A_r x_r \rightarrow A_s z_{k+1,s} = r_{k+1,s}.
\end{equation}
This is equivalent to using the local part of the matrix $A$, $A_s$, on each processor as the preconditioner $M$ in the preconditioned CG,
\begin{equation}
  M=\bigoplus_s A_s,\ s=\mathrm{node\ index}.
\end{equation}
The local nature of $A_s$ makes it possible to perform the preconditioning step concurrently on all the processors without communication. We refer to this as multisplitting preconditioned CG(MSPCG).


\section{Results}

The multisplitting preconditioned CG is applied to solve Dirac equations on three 2+1 flavor lattice ensembles generated with M\"obius domain wall fermions, all with physical input quark masses. Standard CG is used to perform the inversion in the preconditioning step. Instead of adopting a precision based stopping condition, a fixed number of CG iterations, which will be referred as \textit{inner iterations}, are performed for these preconditioning solves. The iterations performed in the overall preconditioned CG will be referred as \textit{outer iterations}. In table \ref{table:result} the numbers of outer iterations needed for the preconditioned CG to converge are reported on the different lattice ensembles, together with the stopping condition for the outer CG(precision) and the processor grid size used. The numbers of iterations to reach the same precision with standard CG are also included for comparison, where the inner iteration number is marked with \textit{plain}.

Typically on these ensembles with $6$ inner iterations the preconditioned CG reduces the outer iteration count by a factor of $3$. More inner iterations reduce the outer iteration count more but the reduction saturates as the inner iteration count increases: with large number of inner iterations the inner CG solves the preconditioning inversion completely and no further numerical benefit can be exploited.

\begin{table}
\renewcommand{\arraystretch}{1.5}
\centering
\begin{tabular}{cccccc}
\hline
\hline
lattice size & $a^{-1}[\mathrm{GeV}]$ & precision & processor grid size & inner iterations & outer iterations \\
\hline
\hline
\multirow{4}{*}{$32^3\times 64$} & \multirow{4}{*}{$1.37$} & \multirow{4}{*}{$10^{-8}$} & $-$ & plain & $13594$ \\
&& & $2^3\times 4$ & $3$ & $9106$ \\
&& & $2^3\times 4$ & $4$ & $6020$ \\
&& & $2^3\times 4$ & $6$ & $5126$ \\
\hline
\hline
\multirow{4}{*}{$64^3\times 128$} & \multirow{4}{*}{$2.36$} & \multirow{4}{*}{$10^{-10}$} & $-$ & plain & $18092$ \\
&&& $4^3\times 8$ & $6$ & $6008$ \\
&&& $4^3\times 8$ & $12$ & $5083$ \\
&&& $4^3\times 8$ & $18$ & $4948$ \\
\hline
\hline
\multirow{2}{*}{$80^2\times96\times 192$} & \multirow{2}{*}{$3.00$} & \multirow{2}{*}{$10^{-10}$} & $-$ & plain & $16783$ \\
&&& $4^2\times 8^2$ & $6$ & $5719$ \\
\hline
\hline
\end{tabular}
\caption{Number of outer iterations need to converge the multisplitting preconditioned  CG for the lattice ensembles tested in this work. \textit{Inner iterations} refers to the fixed number of CG iterations performed for the preconditioning inversion.  Rows marked with \textit{plain} indicate the iteration count for the same standard CG to converge.}\label{table:result}
\end{table}

\section{Conclusion}
Our results show the MSPCG reduces the number of outer iterations needed to solve the DWF Dirac equation, reducing the inter-processor communication at the expense of performing more local inner iterations. 
We observe that executing a fixed number of inner CG iterations for the preconditioning inversion, instead of using a precision based stopping condition, does not jeopardize the convergence of the outer CG. 
This is true even when as few as $3$ inner iterations are performed.
As a consequence the inner iteration count is a parameter that can be tuned to achieve maximum speed up in the trade-off between inter-processor communication and local computation. 

We note that while the multisplitting algorithm can split the general matrix $A$ in a variety of ways, the splitting presented here, used as a preconditioner in CG, makes it equivalent to the additive Schwarz algorithm. (The additive Schwarz algorithm has been used for the Dirac equation inversion for the  fermions\cite{Osaki2010, Babich2011}.) We use the name MSPCG, as it is through the process of applying the multisplitting algorithm to the DWF Dirac equation that we realize the necessity of including the snake terms in the local matrix.

%


\urlstyle{tt}
\printbibliography
%
%
%

\end{document}